\documentclass[twocolumn,aps,showpacs]{revtex4-1}
\usepackage[latin9]{inputenc}
\usepackage{array}
\usepackage{textcomp}
\usepackage{multirow}
\usepackage{amsmath}
\usepackage{amssymb}
\usepackage{graphicx}

\makeatletter

\providecommand{\tabularnewline}{\\}

\usepackage{color}

\def\NOT(#1,#2){\OneQubitGate(#1,#2){$X$}}

\makeatother

\begin{document}
\title{Identification of different silicon vacancy centers in 6$H$-SiC}
\author{Harpreet Singh\textsuperscript{1}, Andrei N. Anisimov\textsuperscript{1,2},
Pavel G. Baranov\textsuperscript{2} and Dieter Suter\textsuperscript{1}\\
 \textsuperscript{1}Fakultät Physik, Technische Universität Dortmund,\\
 D-44221 Dortmund, Germany. \textsuperscript{2}Ioffe Institute, St.
Petersburg 194021, Russia.}
\begin{abstract}
Silicon vacancies in silicon carbide (SiC) have been proposed as interesting
candidates for quantum technology applications such as quantum sensing
and quantum repeaters. SiC exists in many polytypes with different
plane stacking sequences, and in each polytype, the vacancies can
occupy a variety of different lattice sites. In this work, we characterize
and identify the three most important charged silicon vacancies in
the 6$H$-SiC polytype. We record the photoluminescence and continuous-wave
optically detected magnetic resonance spectra at different radio-frequency
power levels and different temperatures. We individually select the
zero-phonon lines of the different silicon vacancies at low temperatures
and record the corresponding optically detected magnetic resonance
(ODMR) spectra. ODMR allows us to correlate optical and magnetic resonance
spectra and thereby resolve a controversy from earlier work.
\end{abstract}
\maketitle

\section{Introduction}

Silicon carbide (SiC) is an interesting material for a range of applications
that rely on its wide bandgap and established manufacturing processes
$\;$\cite{wang-prap-17,rabkowski,izhevskyi2000review,GONZALEZSZWACKI201755,tarasenko-pssb-18,pavunny2021arrays}.
It contains different types of silicon vacancies, carbon vacancies
and silicon-carbon divacancies, which show remarkable spin properties
that make them promising candidates for new quantum technologies.
Silicon vacancies, as a specific example, are negatively charged and
have spin 3/2~\cite{riedel-prl-12}. Irradiation with visible or
near-infrared light results in nonthermal population of the spin states.
Two characteristic parameters of $V_{Si}^{-}$ centers are the emission
wavelength and the zero field splitting.

In the low temperature photoluminescence (PL) spectra, different centers
can be identified by their emission wavelengths. In sequence of increasing
wavelengths, they are therefore labeled $V_{1},$ $V_{2}$ and $V_{3}$.
In terms of the local structure, they are associated with different
lattice sites where the environment has cubic ($k$) or hexagonal
($h$) symmetry. The correspondence between emission wavelength and
lattice sites is a matter of ongoing controversy~\cite{baranov-prb-11,viktor-prb-17,sorman-prb-00,biktagirov-prb-18,davidsson-apl-19}.

In an earlier paper, we characterized silicon vacancies in $6H$-SiC$\;$\cite{breev2022inverted}
in terms of their photoemission as well as their spin Hamiltonian,
using optically detected magnetic resonance (ODMR). Since ODMR correlates
optical properties with energy differences between spin states, it
provides a useful approach for associating the zero-field splittings
with the emission wavelengths. The goal of this work is to identify
the different types of $V_{Si}^{-}$ centers in $6H$-SiC. For this
purpose, we combine photoluminescence (PL) measurements at different
wavelengths and different polarizations with radio-frequency excitation
of the spin transitions to record ODMR spectra under a range of different
conditions, including its temperature dependence.

This work is structured as follows: Section$\;$\ref{sec:System}
introduces the spin system of the sample, as well as its photoluminescence
as a function of the direction and orientation of the emission, and
separates it into the contributions of the 3 types of vacancies. Section
\ref{sec:odmr} shows the temperature-dependent ODMR spectra for emission
parallel and perpendicular to the $c$-axis. To distinguish the contributions
from the different types of centers, we record signals at different
wavelengths separately. Section~\ref{sec:Relaxation-measurements}
contains the spin-lattice and spin-spin relaxation times obtained
by time-resolved ODMR. Section$\;$\ref{sec:conc} contains a brief
discussion and concluding remarks.

\section{System}

\label{sec:System}

For creating silicon vacancies homogeneously, the sample can be irradiated
with neutrons or electrons~\cite{Hain-jap-14,fuchs2015engineering}.
Neutron irradiation leads to more lattice damage than electron irradiation
and creates many other unwanted detects, resulting in faster relaxation
rates of the vacancy spins~\cite{kasper2021engineering}. For this
work, we therefore used electron irradiation of a 6$H$-SiC sample~\cite{singh2021multi}.
More details of the sample preparation are given in Appendix A.

At the silicon vacancy site, four dangling $sp^{3}$ orbitals contribute
four electrons. In addition, the silicon vacancy can capture one or
two electrons, depending on the Fermi level, and become a negatively
charged silicon vacancy ($V_{Si}^{-}$). Here, we only consider centers
with a single negative charge, which have a spin of 3/2~\cite{riedel-prl-12}.
The energy levels of these vacancies can be determined from a PL spectrum.

\subsection{Photoluminescence}

We start the characterisation of the sample with the photoluminescence
spectra. Depending on the transition dipole, whose orientation varies
with the different types of vacancies \cite{breev2022inverted,zhou2021experimental},
the PL emission shows an orientation dependence. We therefore recorded
PL spectra for emission parallel and perpendicular to the c-axis.
The detailed description of the PL setup is given in Appendix~B.

\label{system}

\begin{figure}
\includegraphics[scale=1.1]{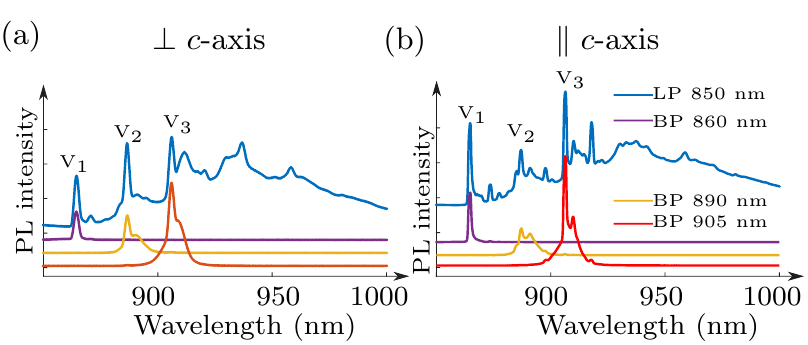}

\caption{Photoluminescence (PL) recorded with different optical filters at
5K (a) for photons emitted $\bot$ to the $c$-axis and (b) $\Vert$
to the $c$-axis. PL spectra obtained by filtering the PL with 860
nm, 890 nm and 905 nm bandpass filters of 10 nm bandwidth to select
the zero-phonon line of $V_{1}$, $V_{2}$, and $V_{3}$ type $V_{Si}^{-}$.}

\label{plodmrfilter}
\end{figure}

Figures$\:$\ref{plodmrfilter}$\:$(a) and (b) show the PL spectra
recorded for emission parallel and perpendicular to the $c$-axis
at 5 K. In addition to the full spectra, where we used an 850 nm long-pass
(LP) filter to suppress scattered laser light, we also recorded spectra
with 860 nm, 890 nm and 905 nm bandpass (BP) filters. The zero phonon
lines (ZPL) of the different $V_{Si}^{-}$ are clearly visible in
the spectra at 865 nm (V$_{1}$), 887 nm (V$_{2}$), and 908 nm ($V_{3}$)$\;$\cite{sorman-prb-00,baranov-prb-11}.
Comparison of the two sets of spectra shows that $V_{1}$ and $V_{3}$
emit more PL parallel to the c-axis, while $V_{2}$ emits more perpendicular
to the $c$-axis~\cite{JANZEN20094354,breev2022inverted,singh-prb-20}.

\begin{figure}
\includegraphics[scale=1.6]{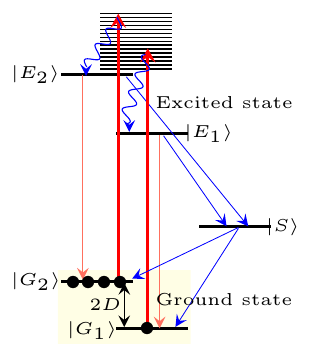}

\caption{Energy level diagram of $V_{Si}^{-}$ and optical pumping scheme.
$\vert G_{i}\rangle$ are the spin states of the electronic ground
state $\vert E_{i}\rangle$, correspond to the electronically excited
state and $\vert S\rangle$ is the shelving state.}

\label{energylevels}
\end{figure}

The assignment of the emission lines to specific lattice sites was
debated in the literature. An early study of $6H$-SiC$\;$\cite{sorman-prb-00},
indicated that $V_{1}$ and $V_{3}$ correspond to $V_{Si}^{-}$ at
the cubic lattice sites $k_{1}$ and $k_{2}$ and $V_{2}$ corresponds
to the hexagonal lattice site $h$. However, more detailed studies
indicate that $V_{1}$ corresponds to the hexagonal lattice site $h$
and $V_{2}$ and $V_{3}$ to the cubic lattice sites $k_{2}$ and
$k_{1}$ respectively$\;$\cite{davidsson-apl-19,biktagirov-prb-18}.
We adopt the latter assignment here.

\subsection{Spin}

We consider the $S=3/2$ spin of a of $V_{Si}^{-}$ center in the
absence of an external magnetic field. Figure$\;$\ref{energylevels}
shows the relevant part of the energy level diagram. The states $\vert G_{1}\rangle$
and $\vert G_{2}\rangle$ are doubly degenerate spin substates of
the electronic ground state with $m_{s}=\pm1/2$ and $\pm3/2$ in
the case of $V_{1}$ and $V_{2}$~\cite{baranov-prb-11,davidsson-apl-19}.
In the case of $V_{3}$, the zero field splitting is negative and
the the lowest energy state is therefore $m_{s}=\pm3/2$~\cite{soltamov-naturecom-19,davidsson-apl-19}.
Similarly, the states $\vert E_{1}\rangle$ and $\vert E_{2}\rangle$
are spin states of the electronically excited state. The state $\vert S\rangle$
represents the shelving state, which has spin $S$=1/2 and plays an
essential role for generating spin polarization by optical pumping$\;$\cite{baranov-prb-11,singh-prb-21}.
Laser illumination excites transitions from the ground states $\vert G_{1}\rangle$
and $\vert G_{2}\rangle$ to the excited states $\vert E_{1}\rangle$
and $\vert E_{2}\rangle$. From the excited states $\vert E_{i}\rangle$,
most of the population falls back to the ground states $\vert G_{i}\rangle$
due to spontaneous emission. A smaller fraction undergoes intersystem
crossing (ISC) to the shelving state $\vert S\rangle$, and from there,
preferentially populates the ground state $\vert G_{2}\rangle$, as
shown in Fig.$\;$\ref{energylevels}.

In the absence of an external magnetic field, the effective spin Hamiltonian
is

\begin{equation}
\mathcal{H}=D(S_{z}^{2}-5/4),\label{hamiltonian}
\end{equation}
where $D$ is the zero-field splitting constant whose value depends
on environment of the center. $S_{z}$ is the electron spin operator
along the $z$ axis ($||$ $c$ axis). In the magnetic resonance spectra,
we therefore expect a single resonance line at frequency $2D$ for
every type of vacancy. Figure \ref{1D_odmr} shows, as an example,
an ODMR spectrum of a 6$H$-SiC sample that contains three different
types of vacancies. It consists of two resonance lines close to 28
MHz and 130 MHz. The assignment of these resonance lines to the different
types of $V_{Si}^{-}$ in $6H$-SiC has been somewhat contentious,
in particular with respect to $V_{1}$. Table~\ref{peak_assigment_table}
provides a summary of data. While Davidsson \emph{et al.}$\;$\cite{davidsson-apl-19}
assume that $V_{1}$ contributes to the resonance near 28 MHz, Biktagirov
\emph{et al.}$\;$\cite{biktagirov-prb-18} suggest that it is close
to zero and therefore not observable. In the following section, we
attempt to resolve this issue by measuring ODMR spectra at low temperatures
and separating the PL from the ZPLs of the 3 different vacancies using
suitable optical bandpass filters.

\begin{widetext}

\begin{table}
\begin{tabular}{|c|c|c|c|c|}
\hline 
\multirow{3}{*}{Center} & Sorman \emph{et al.}$\;$\cite{sorman-prb-00} & \multicolumn{1}{c|}{Davidsson \emph{et al.}$\;$\cite{davidsson-apl-19}} & \multicolumn{1}{c|}{Biktagirov \emph{et al.}$\;$\cite{biktagirov-prb-18}} & \multicolumn{1}{c|}{Astakhov \emph{et al. }\cite{astakhov2016spin}}\tabularnewline
\cline{2-5} \cline{3-5} \cline{4-5} \cline{5-5} 
 & Exp & Exp & Exp & Exp\tabularnewline
\cline{2-5} \cline{3-5} \cline{4-5} \cline{5-5} 
 & $2D^{g}$ (Site) & $2D^{g}$ (Site) & $\vert$$2D^{g}\vert$ (Site) & $\vert$$2D^{e}\vert$ (MHz)\tabularnewline
\hline 
$V_{1}$ & 27.6 MHz ($k_{1}$) & 26.6 MHz ($h$) & 0 ($h$) & 367\tabularnewline
\hline 
$V_{2}$ & 128.4 MHz ($h$) & 128 MHz ($k_{2}$) & 128 MHz ($k_{2}$) & 1030\tabularnewline
\hline 
$V_{3}$ & 27.6 MHz ($k_{2}$) & 27.8 MHz ($k_{1}$) & 28 MHz ($k_{1}$) & 367\tabularnewline
\hline 
\end{tabular}

\caption{Experimental values for the zero field splittings (ZFS) splittings
$2D$ for the 3 types of different $V_{Si}^{-}$ in 6$H$-SiC. The
upper indices $g$ and $e$ refer to the electronic ground and excited
states. The values and the proposed assignments to the different types
of centers are taken from 4 earlier publications.}

\label{peak_assigment_table}
\end{table}

\end{widetext}

\section{Optically detected magnetic resonance}

\label{sec:odmr}

Optically detected magnetic resonance (ODMR) combines optical measurements
with electron spin resonance spectroscopy. Compared to conventional
EPR, this double resonance technique increases the sensitivity and
information content of magnetic resonance$\;$\cite{mr-1-115-2020,carbonera2009optically}.
For low-temperature ODMR, we placed the sample in a liquid helium
flow cryostat and irradiated it with 785 nm laser light. The emitted
PL was collected using a convex lens and focused with another convex
lens on an avalanche photodiode (APD) via a suitable optical filter.
The photocurrent was measured with a lock-in amplifier. A radio-frequency
(RF) field was applied to the sample through a wire (for continuous-Wave
ODMR) or a coil (for pulsed experiments) , terminated with a 50-Ohm
resistor. To reduce background signal, we modulated the RF with a
TTL signal from a digital word generator (DWG) and demodulated the
APD signal with a lock-in amplifier referenced to the TTL signal.
Appendix B provides a detailed description of the ODMR setup.

\subsection{Dependence on RF Power}

\begin{figure}
\includegraphics{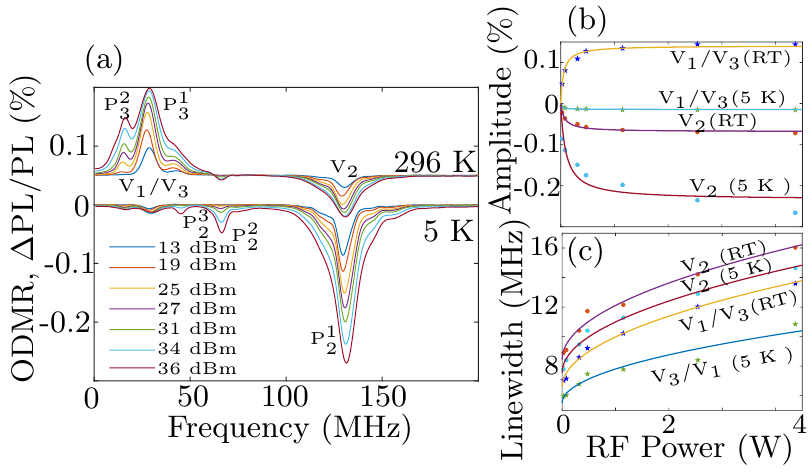}

\caption{(a) ODMR signal vs. frequency recorded with different RF powers in
zero magnetic field at 5 K and 296 K. The horizontal axis is the frequency
in MHz and the vertical axis the relative change of the PL, recorded
by the lock-in amplifier. (b) ODMR signal vs RF power and (c) linewidth
vs RF power.}

\label{1D_odmr}
\end{figure}

Figure~\ref{1D_odmr} (a) shows ODMR spectra recorded with different
RF powers at room temperature (296 K) and at 5 K. These spectra were
recorded with a 850 nm long pass optical filter, which suppresses
the scattered laser light but passes most of the PL from $V_{si}^{-}$.
At low RF power, only two ODMR peaks are visible near 28 MHz and 130
MHz. With increasing RF power, additional resonances appear at 14
MHz ($P_{3}^{2}$), 42 MHz ($P_{2}^{3}$), and 64 MHz ($P_{2}^{2}$).
The peak $P_{3}^{2}$ is due to the absorption of 2 RF photons by
the $V_{1}/V_{3}$ vacancy~\cite{singh2021multi}, whereas the peaks
$P_{2}^{2}$ and $P_{2}^{3}$ are due to 2 and 3 photon absorption
by the $V_{2}$ vacancy~\cite{singh2021multi}. Figure~\ref{1D_odmr}
(b) shows the amplitudes of the 1-photon peaks of $V_{1}/V_{3}$ and
$V_{2}$ vs. the applied RF power: they increase with the applied
power but saturate for $P>0.3$ W. This behaviour can be fitted with
the function
\begin{equation}
S(P)=S_{max}[P/(P_{0}+P)],\label{eq:signalamp}
\end{equation}
where $S(P)$ is the signal amplitude at power $P$, $S_{max}$ is
the asymptotic amplitude and $P_{0}$ the saturation power. The resulting
fitting parameters are given in Table~\ref{odmrfitpar}.

Figure~\ref{1D_odmr} (c) shows the linewidths of the 1-photon peaks
of $V_{1}/V_{3}$ and $V_{2}$ vs. the RF power. The linewidth data
$LW(P)$ were fitted to the function
\begin{equation}
LW(P)=LW_{0}+a\sqrt{P},\label{eq:linewidth}
\end{equation}
where $LW_{0}$ and $a$ are the fitting parameters and their values
are given in Table~\ref{odmrfitpar}.

\begin{widetext}

\begin{table}
\begin{tabular}{|c|c|c|c|c|c|}
\hline 
$V_{si}^{-}$ & Temp & $S_{max}$($\Delta PL/PL$) & $P_{0}$(W) & $LW_{0}$(MHz) & $a$ (MHz W\textsuperscript{-1/2})\tabularnewline
\hline 
\hline 
$V_{1}/V_{3}$ & 5 K & -0.014 $\pm$0.001 & 0.023$\pm$0.012 & 5.5 $\pm$ 0.8 & 2.3 $\pm$ 0.7\tabularnewline
\hline 
$V_{2}$ & 298 K & -0.069 $\pm$ 0.007 & 0.070 $\pm$ 0.043 & 8.4$\pm$0.7 & 3.7$\pm$0.6\tabularnewline
\hline 
$V_{2}$ & 5K & -0.234$\pm$0.054 & 0.097$\pm$0.097 & 7.5 $\pm$ 0.4 & 3.4 $\pm$ 0.4\tabularnewline
\hline 
$V_{1}/V_{3}$ & 298 K & 0.14 $\pm$ 0.01 & 0.05 $\pm$ 0.03 & 6.6 $\pm$ 0.4 & 3.4 $\pm$0.3\tabularnewline
\hline 
\end{tabular}

\caption{Fitting parameters of Eqs~\eqref{eq:signalamp} and \eqref{eq:linewidth}.}

\label{odmrfitpar}
\end{table}

\end{widetext} 

\begin{figure}
\includegraphics[scale=1.13]{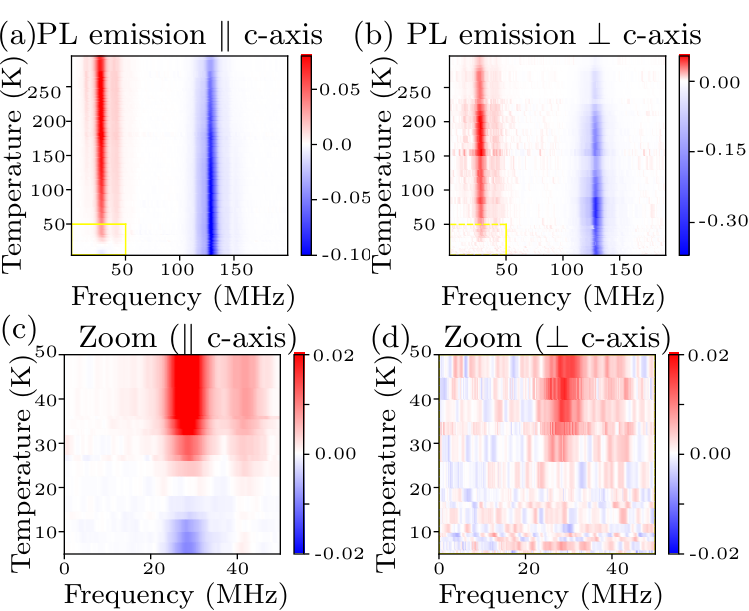}

\caption{ODMR signal as a function of temperature recorded with PL emitted
(a, c) parallel and (b, d) perpendicular to the $c$-axis. The range
from 5 MHz to 50 MHz is shown on an expanded scale in (c, d). The
color scale and the $y$-axis of the 1D spectra are in units of $\Delta$PL/PL\%.}

\label{odmr_temp}
\end{figure}

\begin{figure}
\includegraphics[width=1\columnwidth]{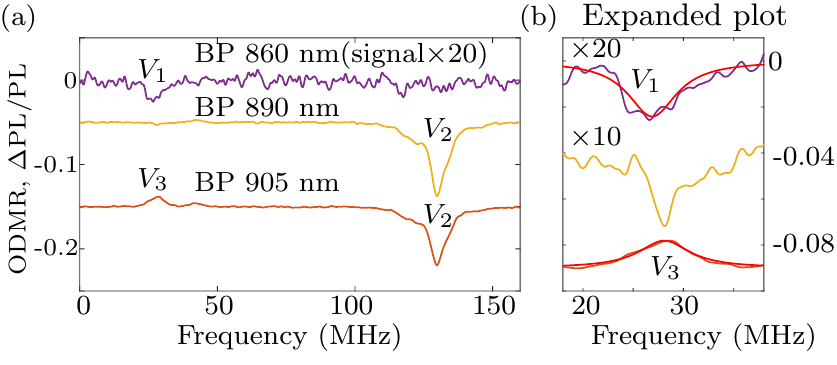}

\caption{(a) ODMR spectra recorded with different optical filters at 28 K for
PL emitted parallel to the $c$-axis. 860 nm, 890 nm and 905 nm bandpass
filters were used to select the zero-phonon lines of $V_{1}$, $V_{2}$,
and $V_{3}$. (b) The range from 18 MHz to 38 MHz, where the $V_{1}$
and $V_{3}$ resonances are located, are shown on an expanded scale.
The main peaks was fitted to a Lorentzian. With the 860 nm and 905
nm BP filters, the center frequencies are 27 MHz and 28.08 MHz, respectively.}

\label{filter30k}
\end{figure}

\subsection{Temperature dependence}

\label{subsec:Temperature-dependence}

Figures$\;$\ref{odmr_temp} (a) and (b) show the ODMR spectra for
the range of temperatures from 5 K to 290 K when the PL is recorded
parallel and perpendicular to the $c$-axis using an RF power of 26
dBm (0.3 W). The horizontal axis corresponds to the RF frequency,
and the vertical axis to the temperature. The color represents the
relative change of PL, as indicated by the scale bar on the right.
The ODMR signal near 130 MHz, which corresponds to $V_{2}$, increases
with decreasing temperature while the ODMR signal near 28 MHz, which
is due to $V_{1}/V_{3}$, decreases. We recorded larger ODMR signals
($\Delta$PL/PL) when the PL is emitted parallel to the $c$-axis
in the case of $V_{1}/V_{3}$, whereas the ODMR signal of $V_{2}$
was larger for emission $\perp c$. In both orientations, the $V_{2}$
signal increases with decreasing temperature, while it gets smaller
in the case of $V_{1}/V_{3}$. However, for PL emitted parallel to
the $c$-axis, the ODMR signal near 28 MHz changes sign. Figure~\ref{odmr_temp}
(c), shows that the ODMR signal vanishes at 22 K and becomes negative
at 15 K. At even lower temperatures, the negative signal increases
in magnitude.

\begin{figure}
\includegraphics{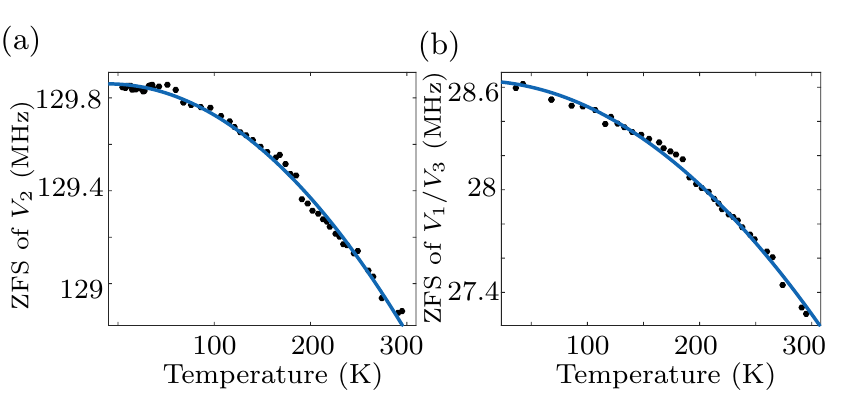}

\caption{Measured temperature dependence of the ZFS of the resonance lines
near 130 (a) and 28 MHz (b).}

\label{zfs_temp}
\end{figure}

Figures.~\ref{zfs_temp} (a) and (b) show the measured frequency
of the ODMR resonance lines that correspond to the ZFS as a function
of temperature. (a) corresponds to the $V_{2}$ center while the resonance
line near 28 MHz is dominated by the signal from $V_{3}$ in the temperature
range from 36 K to 295 K. The extracted data is fitted to the function

\begin{equation}
ZFS(T)=ZFS_{0}+b\,T^{2}.\label{eq:zfstempd}
\end{equation}
The best fits are obtained for the parameters $ZFS_{0}$ = 129.9 $\pm$
0.5 MHz and 28.64 $\pm$ 0.02 MHz , $b$ = (-1.19 $\pm$ 0.02) $\times$
10$^{-5}$ MHz K$^{-2}$ and (-1.52 $\pm$ 0.02) $\times$ 10$^{-5}$
MHz K$^{-2}$ for $V_{2}$ and $V_{1}/V_{3}$, respectively. These
results correspond to an improvement over earlier reports that the
ground-state ZFS does not depend on temperature \cite{astakhov2016spin}.

\subsection{Optically selective detection}

\label{subsec:Optically-selective-detection}

For a better identification of the different ODMR signals, we only
collect PL from specific wavelength ranges by passing it through different
optical filters. Figure$\;$\ref{filter30k} (a) shows the ODMR spectra
recorded with different bandpass (BP) filters at 28 K for PL emitted
$\parallel$ $c$-axis. The top, middle, and bottom spectra were recorded
with 860 nm, 890 nm, and 905 nm BP (10 nm bandwidth) filters, which
predominantly select PL from the zero-phonon line (ZPL) of the $V_{1}$,
$V_{2}$, and $V_{3}$ vacancies, respectively.

The top spectrum (860 nm BP), contains only a small negative peak
at 27 $\text{\ensuremath{\pm}}$ 1 MHz. This differs strongly from
the ODMR spectra in Fig.~\ref{1D_odmr} recorded with the 850 nm
LP filter, where we observed two single-photon resonances near 28
MHz and 130 MHz. Since the 860 nm BP filter selects the ZPL of $V_{1}$,
we assign the negative peak at 27 MHz to $V_{1}$. Since the signal
is quite small under these conditions, we expanded the vertical scale
of the top spectrum by a factor of 20 to make the peak visible.

The middle ODMR spectrum was recorded with a BP filter centered at
the ZPL of $V_{2}$. It is dominated by a signal at 130 $\pm$ 0.01
MHz, which we therefore attribute to $V_{2}$. The bottom ODMR spectrum
was recorded with the 905 nm BP filter; we observe one negative and
two positive peaks. We attribute the positive peak at 28.1 $\pm$
0.6 MHz to the $V_{3}$ center, whose ZPL coincides with the center
of the 905 nm BP filter. The negative peak at 129.93 $\pm$0.03 MHz
is due to the phonon sidebands (PSB) of $V_{2}$, which extend well
into the range of the 905 BP filter.

Figure$\;$\ref{filter30k} (b) shows the frequency range from 18
MHz to 38 MHz, where the signals from $V_{1}$ and $V_{3}$ are centered,
on an expanded scale. In the top spectrum, the signal from $V_{1}$
dominates, in the bottom spectrum that from $V_{3}$.

As an additional way to distinguish between the PL emitted by $V_{1}$
and $V_{3}$, we also measured ODMR spectra with polarizers at different
orientations. The corresponding data are presented in appendices
E and F.

\subsection{Separating the contributions from $V_{1}$ and $V_{3}$}

As discussed in the introduction, there is conflicting evidence on
the ZFS of $V_{1}$ and $V_{3}$. The experiments presented in subsections
\ref{subsec:Temperature-dependence} and \ref{subsec:Optically-selective-detection}
were designed to provide data for distinguishing between them. The
results can be interpreted consistently if we assume that the ZFS
splittings of $V_{1}$ and $V_{3}$ are both close to 28 MHz but have
slightly different frequencies (27 MHz and 28 MHz) and contribute
to the ODMR spectrum with opposite signs : for $V_{1},$ the ODMR
signal is negative ($\Delta PL/PL$\textless{} 0), for $V_{3}$ it
is positive. Since the combined linewidth is larger than the separation
between the two resonance frequencies, only a single line is obersved
and the two signals partially cancel. Since the PL from the two centers
show different polarisation, the positive and negative signal contribution
depend differently on the orientation of the polarizer, as shown in
Appendix F.

\begin{figure}
\includegraphics{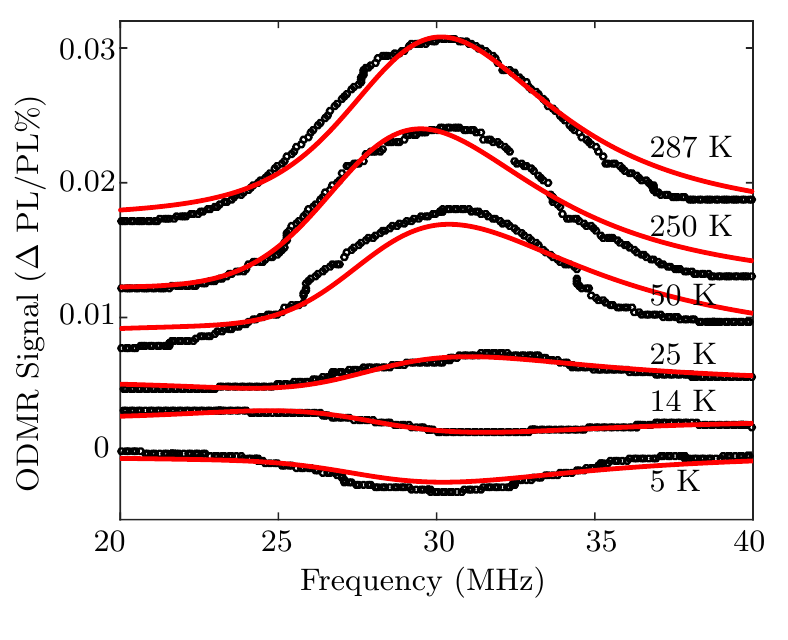}

\caption{ODMR spectra measured with the 850 nm LP filter at different temperatures
for PL emitted parallel to the $c$-axis. Only the region of the $V_{1}$
and $V_{3}$ resonances are shown and the spectra are fitted to a
sum of two Lorentzians.}
\label{odmr_tempfitting}
\end{figure}

The signals from the two centers also have different temperature dependence.
At low temperatures, the $V_{1}$ signal at 27 MHz dominates, resulting
in an overall negative signal but at high temperatures, the $V_{3}$
signal at 28 MHz dominates, resulting in an overall positive signal.
For a more detailed analysis, we compare the experimental signal at
different temperatures to a theoretical spectrum consisting of two
Lorentzians with the frequencies of $V_{1}$ and $V_{3}$ and opposite
amplitudes

\begin{equation}
\ensuremath{\ensuremath{S(f,T)=\frac{A_{1}(T)\,\sigma_{1}}{(f-ZFS_{1})^{2}+\sigma_{1}^{2}}+\frac{A_{3}(T)\,\sigma_{3}}{(f-ZFS_{3})^{2}+\sigma_{3}^{2}}}}.\label{eq:double_lorz}
\end{equation}

The frequencies $ZFS_{i=1,3}$=$2D$ were calculated from Eq.~\eqref{eq:zfstempd}
for different temperatures, using the temperature coefficient $b$
= (-1.52 $\pm$ 0.02) $\times$ 10$^{-5}$ MHz K$^{-2}$ for $V_{1}$
and $V_{3}$. The values of the ZFS at $T=0$ were determined from
the spectra measured with optical filters at 28 K as $ZFS_{0}$ =
26.99 MHz and 27.99 MHz for $V_{1}$ and $V_{3}$ . The linewidths
$\sigma_{1}=5.9$ MHz and $\sigma_{3}=6.1$ MHz were also taken from
the spectra measured at 28 K.

Figure~\ref{odmr_tempfitting} shows the ODMR spectra for the frequency
range 10 to 50 MHz at different temperatures, together with a fitted
signal, where only the amplitudes $A_{i}$ were taken as fitting parameters
to measure the contributions from $V_{1}$ and $V_{3}$. Overall,
the superposition of the two resonance lines yields excellent agreement
with the experimental data. 

For PL emitted $\perp c$, the signal near 28 MHz decreases with temperature
but does not become negative, as shown in Fig.~\ref{odmr_temp}~(d).
This is also consistent with the assumption that the negative signal
is associated with $V_{1}$: as shown in appendix E, $V_{1}$ emits
less PL perpendicular to the $c$-axis as shown in Fig.\ref{plodmrfilter}
and therefore can not dominate over the contribution from $V_{3}$.

\section{Relaxation measurements}

\label{sec:Relaxation-measurements}

In this section we focus on the evolution of the spin system towards
thermal equilibrium, which is driven by interactions with the environment~\cite{abragam-book}.
Each type of silicon vacancies has a different structure and therefore
different interactions with the environment that can be probed by
relaxation measurements.

\subsection{Population relaxation}

\begin{figure}
\includegraphics{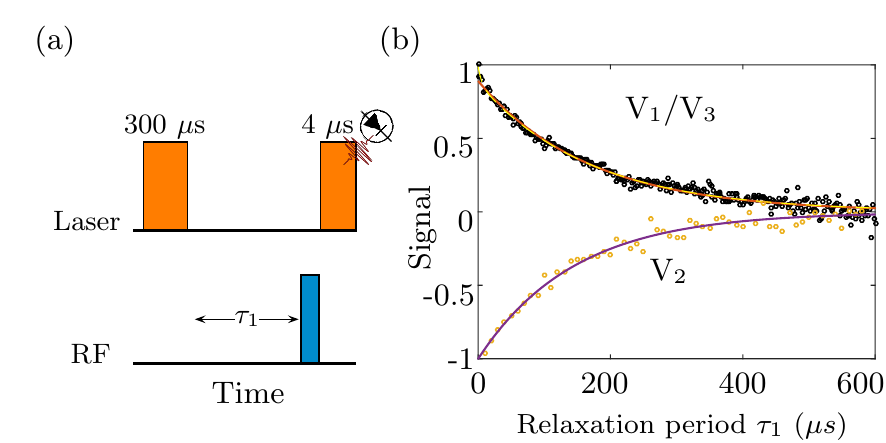}

\caption{(a) Pulse sequence used to measure the $T_{1}$ relaxation. The orange
and blue rectangles represent the laser and RF pulses. The duration
of the laser pulses is written above each pulse. (b) Resulting signal
(normalized) as a function of the delay $\tau_{1}$, measured at room
temperature. The experimental data (circles) for $V_{2}$ and $V_{1}$/$V_{3}$
are fitted to function Eq.\ref{eq:single}.}

\label{t1fig}
\end{figure}

The spin-lattice interaction drives the spin system towards thermal
equilibrium with its environment. Under our experimental conditions,
the thermal equilibrium spin state is completely unpolarised, i.e.
the density operator is proportional to the unit operator. The time
scale on which the system approaches this equilibrium state is characterized
by the spin-lattice relaxation time $T_{1}$. Figure~\ref{t1fig}(a)
shows the pulse sequence used to measure the $T_{1}$ relaxation time.
A 300 $\mu$s laser pulse was applied to polarize the spin system,
which was then allowed to evolve for a time $\tau_{1}$. An RF pulse
with flip angle $\pi$ was applied before measuring the remaining
spin polarisation with a second laser pulse of 4 $\mu$s. The measured
signal was subtracted from the reference signal obtained without applying
the RF pulse\cite{singh-prb-20,singh-prb-21,singh2021multi}. Figure~\ref{t1fig}(b)
shows the measured signal vs. the delay $\tau_{1}$. The recorded
experimental data can be fitted by an exponential

\begin{equation}
S_{\pi}(\tau_{1})-S_{0}(\tau_{1})=A~e^{-\tau_{1}/T_{1}}.\label{eq:single}
\end{equation}
At room temperature, the extracted relaxation times $T_{1}$ were
145$\pm$2 $\mu$s for $V_{2}$ and 166 $\pm$ 2 $\mu$s for $V_{1}/V_{3}$.

\subsection{Dephasing}

Relaxation affects not only the populations, but also the coherence
of the spins. While population relaxation is dominated by fluctuations
at the Larmor frequency, coherence dephases under the influence of
(quasi-) static interactions. A free induction decay (FID) for both
$V_{Si}^{-}$ was measured using the Ramsey scheme$\;$\cite{ramsey-pr-50}.
The dephasing time $T_{2}^{*}$ was 42 $\pm$4 ns for $V_{2}$ and
65 $\pm$ 5 ns for $V_{1}/V_{3}$~\cite{singh2021multi}. Details
of the pulse sequence used and experimental data obtained are given
in Appendix D.

\begin{figure}
\includegraphics{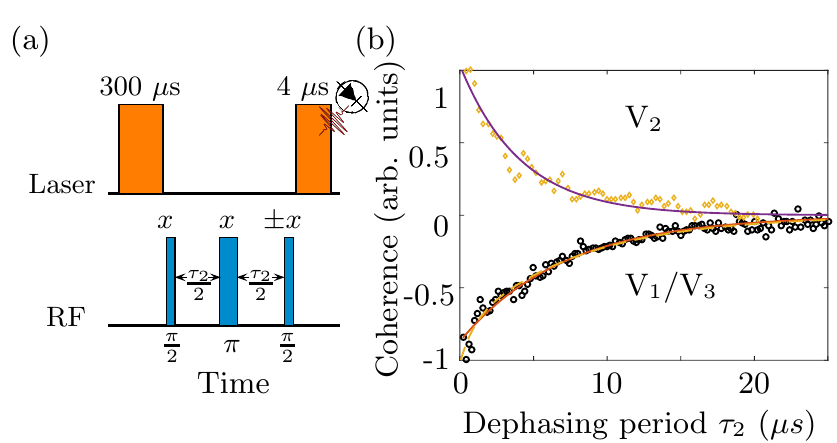}

\caption{(a) Pulse sequence for measuring the spin coherence time with a refocusing
pulse. The red and blues rectangles represent the laser and RF pulses,
respectively. (b) Decay of the spin coherence during the echo sequence.
The experimental data (circles) for $V_{2}$ and $V_{1}/V_{3}$ are
fitted to exponential functions.}

\label{spinecho}
\end{figure}

Next, we focus on homogeneous dephasing, which can be measured using
the spin-echo experiment~\cite{PhysRev.80.580}. We refer to the
time constant of the spin-echo decay as $T_{2}$. Figure~ \ref{spinecho}~(a)
shows the pulse sequence. The system was again polarised with a 300
$\mu$s laser pulse. The first RF pulse with flip-angle $\pi/2$ generated
coherence between the states $\pm3/2\longleftrightarrow\pm1/2$. The
inhomogeneous dephasing was reversed with a $\pi$ pulse, which generated
a spin-echo at time $\tau_{2}$ after the initial $\pi/2$ pulse.
A second $\pi/2$ RF pulse at the time of the echo converted the remaining
coherence into population difference, which was measured by the final
laser pulse~\cite{singh-prb-20,singh-prb-21}. Figure~\ref{spinecho}(b)
shows the experimental signal vs. the delay $\tau_{2}$. The recorded
experimental data can again be fitted to a single exponential and
the extracted dephasing times $T_{2}$ were 4.1 $\pm$ 0.4 $\mu$s
and 7.1 $\pm$1.7 $\mu$s for $V_{2}$ and $V_{1}/V_{3}$, respectively,
roughly 2 orders of magnitude longer than $T_{2}^{*}$. The spin-echo
data for $V_{1}/V_{3}$ does not fit well to the single exponential,
so we also compared it to a sum of 2 exponentials $(A~e^{-\tau_{2}/T_{2}^{s}}+B~e^{-\tau_{2}/T_{2}^{f}}$),
as shown in Fig. ~\ref{spinecho} with a yellow dashed curve ( $A$=
-0.8 $\pm$0.1, $T_{2}^{s}$= 7.7 $\pm$ 0.3 $\mu$s, $B$= -0.2 $\pm$
0.1, and $T_{2}^{f}$= 0.8 $\pm$ 0.4 $\mu$s.).

Next, we applied multiple refocusing pulses to further extend the
coherence lifetime. These refocusing pulses remove the dephasing effect
of non-static perturbations that have longer correlation times than
the spacing between the refocusing pulses. The measured dephasing
time $T_{2}^{CPMG}$ for the $V_{2}$ ($V_{1}$/$V_{3}$) type of
$V_{Si}^{-}$ spin ensemble was 47 $\pm$ 7 $\mu$s (59 $\pm$ 3 $\mu$s)
when the spacing between the refocusing pulses was 200 ns (100 ns),
and the duration of each $\pi$ pulse was 38 ns (40 ns). Details of
the pulse sequence used and plots of the experimental data are given
in Appendix D.

\begin{table}
\begin{centering}
$\begin{tabular}{|c|c|c|c|c|}
\hline  Exp  & \multicolumn{2}{|c|}{\ensuremath{^{13}}C (1.1\ensuremath{\%}) \&\ensuremath{^{29}}Si(4.7\ensuremath{\%})  } & \multicolumn{2}{|c|}{ \ensuremath{^{13}}C (4.7\ensuremath{\%}) \& \ensuremath{^{29}}Si(0.1\ensuremath{\%}) \cite{singh-prb-20}}\\
\hline   &  V\ensuremath{_{1}}/V\ensuremath{_{3}}  &  V\ensuremath{_{2}}  &  V\ensuremath{_{1}}/V\ensuremath{_{3}}  &  V\ensuremath{_{2}} \\
\hline  T\ensuremath{_{1}}  &  166\ensuremath{\mu}s  &  142\ensuremath{\mu}s  &  142\ensuremath{\mu}s  &  107 \ensuremath{\mu}s \\
\hline  T\ensuremath{_{2}^{*}}  &  65\,ns  &  42\,ns  &  38\,ns  &  31\,ns \\
\hline  T\ensuremath{_{2}^{SE}}  &  7.1\ensuremath{\mu}s  &  4.6\ensuremath{\mu}s  &  3.7\ensuremath{\mu}s  &  3.3\ensuremath{\mu}s \\
\hline  T\ensuremath{_{2}^{CPMG}}  &  57\ensuremath{\mu}s  &  47\ensuremath{\mu}s  &  56\ensuremath{\mu}s  &  51\ensuremath{\mu}s 
\\\hline \end{tabular}\;$
\par\end{centering}
\caption{Comparison of room-temperature relaxation times of two 6H-SiC samples.}

\label{comp_sample}
\end{table}

Table~\ref{comp_sample}, shows the comparison of the relaxation
times measured in our previous paper where the 6H-SiC sample contained
a higher concentration of $^{13}$C but lower concentration of $^{29}$Si~\cite{singh-prb-20}.
In the present sample, $T_{1}$ is slightly longer than in the sample
used in \cite{singh-prb-20}, which indicates a somewhat lower concentration
of paramagnetic centers~\cite{simin-prb-17}. The longer dephasing
times $T_{2}^{*},$ $T_{2}$ can be attributed to the lower concentration
of $^{13}$C, which represent the nearest neighbors (NN) of the silicon
vacancy. The higher concentration of $^{29}$Si does not have a significant
effect, since the silicon atoms have a much weaker hyperfine interaction.

\section{Discussion and Conclusion}

\label{sec:conc}

The $V_{Si}^{-}$ center in SiC has attracted considerable interest
in the context of emerging quantum technologies, since its spin can
be optically polarized, has a sufficiently long coherence time and
can be coherently controlled. These properties make it an attractive
candidate for quantum information and quantum sensing. To develop
this potential, it is essential to more precisely characterise its
properties. They depend, e.g., on the lattice site at which the vacancy
is created. In the $6H$-SiC polytype, three different lattice sites
exist, and the corresponding vacancies are referred to as $V_{1}$,
$V_{2}$ and $V_{3}$.

In this work, we have gathered new data on all three types of vacancies.
In particular, we related the properties of the PL of each center
to the magnetic resonance by performing ODMR experiments for different
optical frequencies, different optical polarizations and different
directions of PL emission (parallel and perpendicular to the $c$-axis
of the sample). We found that the $V_{1}$ and $V_{3}$ vacancies
emit more PL parallel to the $c$-axis, whereas $V_{2}$ emits more
perpendicular to the $c$-axis. ODMR experiments detecting photons
that are emitted mostly by one type of center allowed us to determine
the zero-field splitting of the $V_{1}$ vacancy, for which different
values had been reported in previous works~\cite{biktagirov-prb-18,davidsson-apl-19}.
Our results agree with those of Ref:~\cite{davidsson-apl-19}. With
the identification of the ODMR frequencies of all three vacancies,
we can explain the temperature dependence and change of sign of the
ODMR signal near 28 MHz: At the high temperature, the signal from
$V_{3}$ dominates but at low temperature the signal from $V_{1}$
dominates, which is negative. In real systems, the ZFS typically shows
a small temperature variation. We checked the ZFS variation with temperature
and found that the values increase by about 1 MHz as the temperature
decreases from room temperature to 5 K.

Further, we measured the transverse and longitudinal relaxation rates.
The relaxation rates measured here were lower than those reported
earlier in a different sample~\cite{singh-prb-20} that contained
a higher concentration of $^{13}$C. In the spin-echo decay for $V_{1}/V_{3}$,
does not fit well to a single exponential, which may indicate differences
in the relaxation of $V_{1}$ and $V_{3}$ . In future work, we will
try to obtain more specific data for the relaxation times of the different
centers and determine the different hyperfine coupling constants.

\section*{Appendix A: Sample}

Synthesized SiC crystals were grown with a low content of background
impurities. We synthesized polycrystalline sources from semiconductor
silicon and spectrally pure carbon. Spectrally pure graphite in powder
form and polycrystalline silicon was chosen as sources. They were
degassed using a resistive heating growth machine over 2 hours at
2200° C and 10\textsuperscript{-3} Torr under vacuum to prepare the
crucible and internal furnace reinforcement for synthesis. After that,
a stoichiometric mixture of carbon and silicon powders was loaded
into the crucible. Manufacturing silicon carbide powder took place
in a vacuum. Graphite with a minimum of background impurities was
used to make the crucible, such as Mersen 6516PT. It should note that
all parts of the crucible should be made from the same graphite. To
avoid the destruction of parts, when the crucible is heated above
the synthesis temperature of the source (above 1600°C). We used silicon
carbide crystals polytype 6$H$ as seeds. To grow crystals, the following
conditions were used: 2050 ° C, 1 to 5 Torr argon pressure, 99.9999\%
pure argon, a growth rate of 150 $\mu$m / h. Crystals that grew under
these conditions had no micropores more than 3 cm\textsuperscript{-2}.

\section*{Appendix B: PL setup}

For measuring the photoluminescence (PL) at different temperature,
the sample was cooled down using a helium cryostat. A tunable (799-813
nm) single-mode diode laser DL pro from Toptica Photonics modulated
by a chopper was used to excite the sample optically. With the help
of two convex lenses, PL was collected parallel to the c-axis of sample.
The collected PL passed through a monochromator (Spex 1704) via an
850 nm long-pass filter (Thorlabs). For detecting the PL, an avalanche
photodiode(APD) module with a frequency bandwidth from dc to 100 kHz
(C5460-1 series from Hamamatsu) was attached to the monochromator.
The output voltage of the APD was measured with the lock-in amplifier.
The sync signal from the chopper was used as a reference for the lock-in
amplifier$\;$\cite{singh-prb-20}.

\section*{Appendix C: ODMR setup}

\begin{figure}
\includegraphics[scale=1.1]{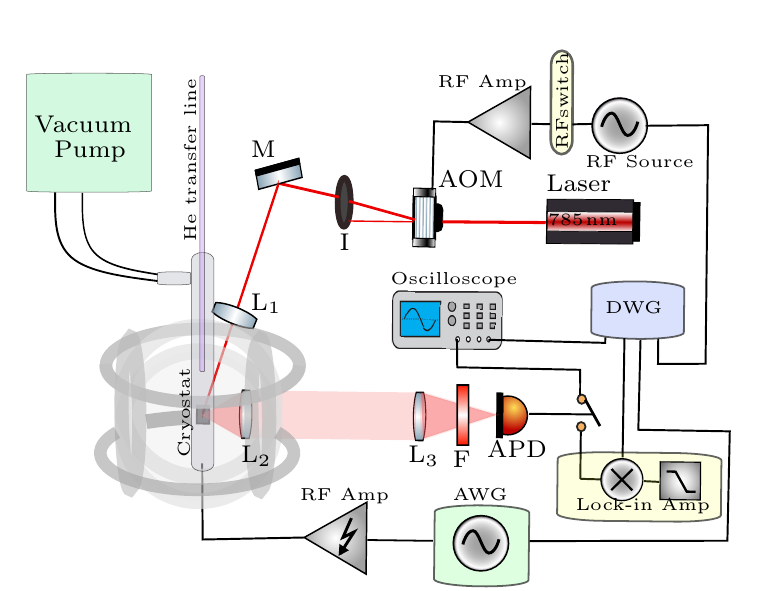}

\caption{Experimental setup for measuring ODMR at low temp.}

\label{odmr_setup}
\end{figure}

Figure~\ref{odmr_setup} shows the setup used for the cw- and time-resolved
ODMR measurements. For cooling the sample, we used a liquid helium
flow cryostat (MicrostatHe-R from Oxford instruments). We used a turbopump
for creating a vacuum in the cryostat pressure less than $10^{-6}$
mbar (from Pfeiffer vacuum). An LD785-SE400 diode laser was used as
the light source (a laser diode controller (LDC202C series) and temperature
controller (TED 200C). The laser light pulses were generated by an
acoustic-optical modulator (NEC model OD8813A). We used three orthogonal
coil pairs to apply a static magnetic field to the sample. APD module
(C12703 series from Hamamatsu) was used to record the PL signal. APD
signal was recorded with the PicoScope 2000 series USB oscilloscope
card during pulse mode ODMR experiments. For cw-ODMR, the signal from
APD was recorded with the lock-in (SRS model SR830 DSP). Analog Devices'
AD9915 direct digital synthesizer (DDS) was used as an RF source for
cw-ODMR experiments. We used a Hunter Micro DAx14000 arbitrary wave
generator (AWG) for pulsed ODMR experiments. An RF signal from the
source was amplified with an RF amplifier (LZY-22+ from mini circuits).
For feeding RF power to a sample, we used wire and coil terminated
with a 50-ohm resistor for continuous-wave and pulsed ODMR experiments,
respectively. We generated TTL (transistor transistor logic) pulses
using a digital word generator (DWG; SpinCore PulseBlaster ESR-PRO
PCI card) to trigger the laser RF pulses.

\section*{Appendix D: ZFS with temperature}

\section*{Appendix D: FID and CPMG}

\begin{figure}
\includegraphics{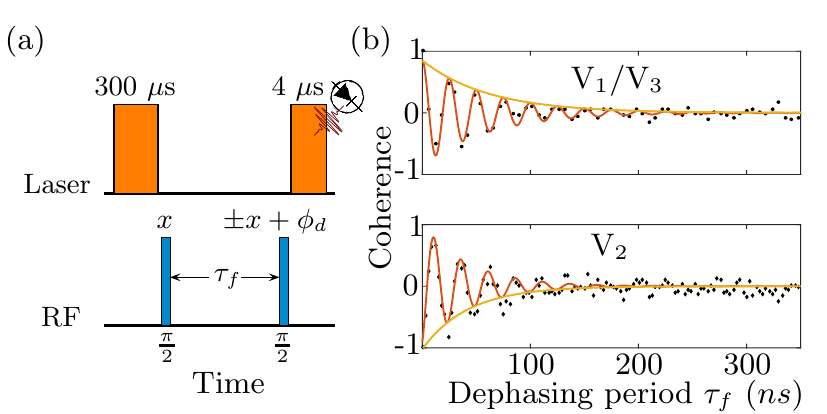}

\caption{(a)Pulse sequence to measure the FID of $V_{Si}^{-}$. (b) The experimental
recorded FID signal of $V_{Si}^{-}$ at room temperature.}

\label{nfid}
\end{figure}

Figure.$\;$\ref{nfid} (a) shows the pulse sequence for the FID measurement.
A laser pulse of 300 $\mu$s and a power of 100 mW was used to polarize
the vacancy spin ensemble. A delay $\tau_{f}$ was given between the
two $\pi/2$ RF pulses, and the PL signal was using the second laser
pulse. The phase of the first $\pi/2$ pulse was along x. For the
second $\pi/2$ pulse, it was along x+$\phi_{d}$. We repeated the
same sequence for the reference scan except for the phase of the second
$\pi/2$ pulse was $-x+\phi_{d}$. Here the $\phi_{d}=\nu_{det}\tau_{f}$
and FIDs measured with the detuning frequency $\nu_{det}$= 40 MHz.
Figure.$\;$\ref{nfid} (b) shows the signals recorded for the both
$V_{Si}^{-}$ as a function of dephasing time delay $\tau_{f}$. The
experimental recorded signal of $V_{2}$ ($V_{3}$) vacancy fitted
to the function

\begin{equation}
S_{x+\phi_{d}}-S_{-x+\phi_{d}}=A\;cos(2\pi\nu_{det}\tau_{f}+c)e^{-\tau_{f}/T_{2}^{*}},\label{eq:fid}
\end{equation}

where $S_{\pm x+\phi_{d}}$ is the average PL signal measured with
$\pi/2$ RF detection pulse of phase $\pm x+\phi_{d}$. After fitting
the experimental signal in the Eq.\ref{eq:fid}, the $T_{2}^{*}$=
42 ns (65$\pm$5 ns).

\begin{figure}
\includegraphics{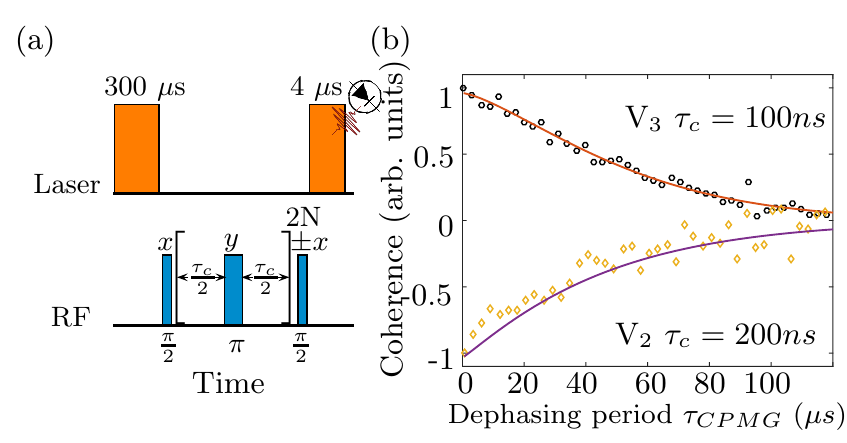}\caption{(a) Pulse sequence for measuring the spin coherence time under the
CPMG sequence. (b) The experimental recorded decay of the spin coherence
of $V_{Si}^{-}$ during the CPMG sequence at room temperature.}

\label{cpmg}
\end{figure}

Figure$\;$\ref{cpmg}(a) shows the pulse sequence for measuring the
spin coherence time under the CPMG sequence. We first polarized the
spin ensemble with the 300 $\mu$s laser pulse, and then the coherence
was created with the $\pi$/2 pulse. The CPMG sequence was applied
i.e., a train of $2N$ $\pi$ pulses applied along the $y$-axis,
which was separated by the delay $\tau_{c}$/2 from each other. Atlast
coherence was coverted into the population with the application of
second $\pi$/2 pulse along the $x$-axis and PL signal was measured
with the second laser pulse. The experiment was repeated for recording
the reference signal for which the pulse sequence was same except
the second $\pi$/2 pulse was applied along the $-x$-axis. The difference
between the experimentally measured PL signals was fitted to the function

\begin{equation}
S_{x}-S_{-x}=A\;e^{-(\tau_{CPMG}/T_{2}^{CPMG})^{n}},\label{eq:cpmg}
\end{equation}

where the time period $\tau_{CMPG}$=2N($\tau_{c}+\pi$ pulse duration),
2N is number of $\pi$ pulses.

\section*{Appendix E: PL Polarization}

\begin{figure}
\includegraphics[scale=1.1]{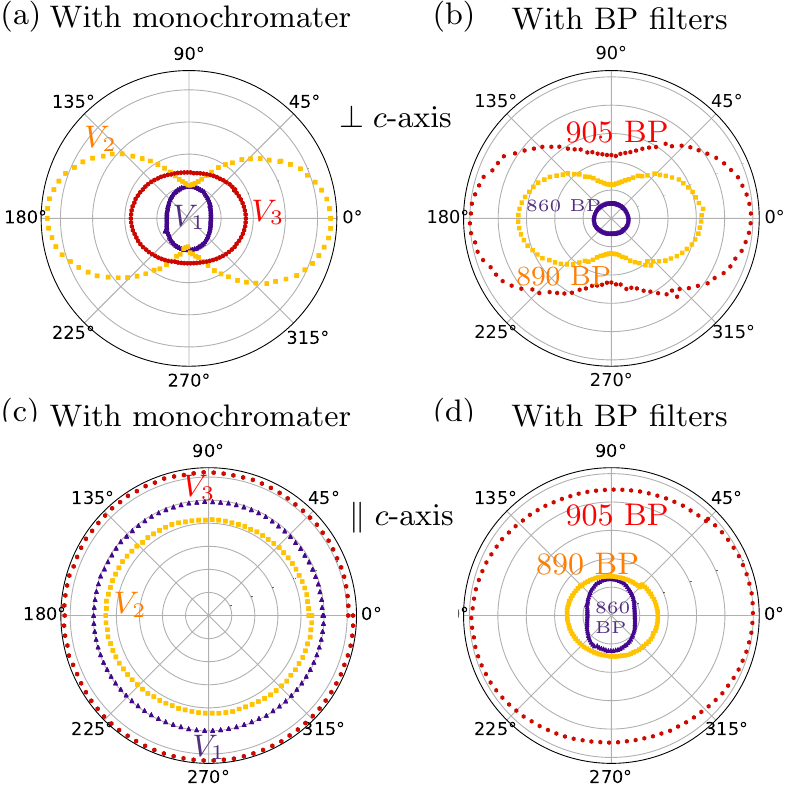}

\caption{Polarization plots of the ZPL intensities for the different centers
at 5 K, (a) with monochromator, and (b) with bandpass filters for
emission perpendicular  to the $c$-axis ; (c) with monochromator,
and (d) with bandpass filters, for emission parallel to the $c$-axis.
$\theta$ is the angle difference between the initial and final position
of the polarizer's transmission axis and in case of $\perp$ $c$-axis
orientation, $0^{\circ}$ is $\parallel$ to $c$-axis.}

\label{polar}
\end{figure}

Figure$\;$\ref{polar} shows the polarization plots of the $V_{1}$,
$V_{2}$ and $V_{3}$ ZPL intensities recorded perpendicular and parallel
to the $c$-axis at 5K. A polarizer was used for selecting the particular
PL polarization. Before detecting the signal, individual ZPLs of the
different silicon vacancies were selected using the monochromator
or suitable bandpass filters. With the monochromator, we can select
a very narrow bandwidth and minimise the contribution from the phonon
sidebands of the other vacancies. With the optical bandpass filters,
the width of the passband is larger (10 nm), but it has the advantage
of passing more signal, which is important for the ODMR measurements.
Figure$\;$\ref{polar} (a) shows the data for PL emission perpendicular
to the c-axis using the monochromator. The emission of $V_{2}$ is
linearly polarized perpendicular to the c-axis, while that of $V_{1}$
is almost unpolarized$\;$\cite{breev2022inverted}. In the case of
$V_{3}$, a significant signal contribution comes from the phonon
side band (PSB) of $V_{1}$ and $V_{2}$. The data in figure$\;$\ref{polar}
(b) were recorded with different bandpass filters; accordingly, they
contain larger contributions from PSB other centers, which results
in lower polarization. In contrast to previous work~\cite{breev2022inverted}
we did not apply a background correction, since the available data
do not allow a reliable separation into contributions from the different
ZPLs and sidebands.

\section*{Appendix F: ODMR Polarization}

\begin{figure}
\includegraphics{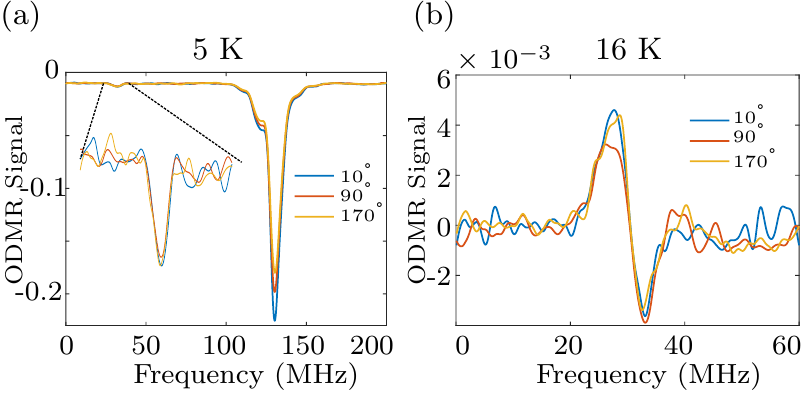}

\caption{(a) Plot of ODMR recorded for different polarizations at 5 K. (b)
ODMR of $V_{1}/V_{3}$ for different polarizationss at 16 K. In both
cases, PL was emitted $\Vert$ to the $c$-axis. $\theta=0$ corresponds
to vertical polarization.}

\label{pol_odmr}
\end{figure}

Since we were not able to separate the ODMR signal of $V_{1}$ in
PL emission $\bot$ $c$-axis we recorded ODMR spectra for different
orientations of the polarization for the PL emitted $\Vert$ $c$-axis.
Figure$\;$\ref{pol_odmr}~(a) shows several ODMR spectra recorded
at 5 K for different orientations of a polarizer in the detection
path The amplitude of the signal near 28 MHz does not change significantly,
but the signal near 130 MHz varies by $\approx$ 21 \%. Figure$\;$\ref{pol_odmr}~(b)
shows the ODMR spectrum measured at 16 K from 0 to 60 MHz. At this
temperature, the signal appears to consist of a positive and a negative
contribution. The amplitude of the positive peak varies with the polarisation
by $\approx31$ \% while the negative peak remains constant. This
appears to be consistent with the different polarization dependence
of $V_{1}$ and $V_{3}$ discussed in Appendix E.
\begin{acknowledgments}
This work was supported by the Deutsche Forschungsgemeinschaft in
the frame of the ICRC TRR 160 (Project No. C7)and by RFBR, project
number 19-52-12058.
\end{acknowledgments}

\section*{Note}

All experiments were done on the sample provided by A. N. A. and P.
G. B well before the EU sanctions. H.S and D.S did experimental measurements
and data analysis. To our knowledge, these results do not have any
short-term economic or military relevance.

\end{document}